# Characterisation of polystyrene coatings after plasma immersion ion implantation and adsorption of protein

S Dekker[1], A Kondyurin[1], B Steel[1], M M M Bilek[1], D R McKenzie[1] and M James[2,3]

[1]Applied and Plasma Physics (A28), The University of Sydney, New South Wales 2006, Australia

[2]The Bragg Institute, Australian Nuclear Science and Technology Organisation (ANSTO), New Illawarra Road, Menai, New South Wales 2234, Australia

[3]School of Chemistry, The University of New South Wales, Sydney, NSW, 2052, Australia

E-mail: kond@mailcity.com

**Abstract**

A polystyrene film spun onto polished silicon substrates was implanted with either nitrogen or argon ions using plasma immersion ion implantation (PIII) and subsequently investigated by X-ray and neutron reflectometry, UV-VIS and FTIR ellipsometry, as well as by FTIR and Raman spectroscopy. The depth profile of the densified carbon structures resulting from the ion collision cascades in the polystyrene coating are clearly observed by both X-ray and neutron reflectometry. Argon ions produce a higher density modified layer at a shallower depth than nitrogen ions. The thickness measured for these graded layers agrees with the expected depths of ion implantation as calculated by SRIM. The sensitivity of X-ray and neutron reflectometry allows resolution of density and hydrogen content gradients within the graphitized layers. The treated layers were found to covalently immobilized protein directly from solution. The tropoelastin protein monolayers immobilized on the surface were characterized. Tropoelastin remained on the surface after SDS washing. PACS 52.77.Dq

**Keywords**: Plasma immersion ion implantation, X-ray reflectometry, neutron reflectometry, ellipsometry, protein, polystyrene

# 1. Introduction

Ion beam implantation is a versatile technique for the modification of polymer surfaces [1-3]. Plasma immersion ion implantation (PIII) is a technique in which a high voltage bias potential is applied to a substrate immersed in a plasma [4]. Ions crossing the high voltage sheath formed around the substrate are accelerated as they approach the substrate. If the voltage is high enough, the ions will be implanted below the surface of the substrate, restructuring the material. The interactions between the target atoms and the incoming ions lead to displacement and excitation of atoms and electrons along the ions path. The interactions cause chemical and structural changes such as cross-linking between polymer macromolecules, scission of the polymer backbone, degassing of volatile side groups or oxidation of the surface upon exposure to the atmosphere. Due to variation of ion energy along the path the modifications are strongly varying with distance from the surface. Layers closest to the surface are etched by the plasma, while sub-surface layers are graphitized due to cross-linking and volatile side groups are removed.

Polymers modified using PIII have found applications in the aeronautics [5], microelectronics [6] and in medicine [7]. Two promising future applications are functionalization of substrates for protein based biosensors, particularly in light of recent evidence that the process can increase the density of protein attachment and the longevity of bioactivity of attached proteins on polymeric materials [8-10].

In many applications, the mechanical integrity of the modified layer is important. Although having significant differences, a previous study by Koval [11] of ion beam etched polymethylmethacrylate (PMMA) thin films on Si using low energy Ar ions (250 eV) revealed a sequence of strata within the polymer. The presence of such strata may have the effect of compromising strength and durability. The depth of the modified layer has implications for the covalent coupling capability. It is therefore important to precisely determine the structure and composition as a function of depth in PIII modified films.

X-ray and neutron reflectometry are important techniques for the investigation of nanoscale thin films and adsorbed layers on surfaces [12-17]. Both techniques have the capacity to non-destructively probe the composition and structure of buried layers within films with essentially atomic resolution. Recently, X-ray reflectometry has been used for studies of protein and cell adsorption on various surfaces including plasma treated surfaces [18-24]. In this paper, we employ X-ray and neutron reflectometry in combination with UV-VIS and FTIR ellipsometry and FTIR and Raman spectroscopy to gain an understanding of the subsurface structure in PIII modified polystyrene and to investigate adsorbed protein layers on the modified surfaces.

# 2. Materials and Methods

*2.1 Sample preparation*

Polystyrene films were spin coated onto (100) polished silicon wafers (1 mm thickness × 100 mm diameter) to a nominal thickness of 15, 25 and 100 nm. The polymer solution consisted of polystyrene (Austrex 400 from Polystyrene Australia Pty. Ltd.) in HPLC toluene (Labscan, purity > 99.9%) concentrations of 4, 6 and 20 g/l. The wafers were spun at a rate of 3250 rpm for a total time of 20.0 s. The first 1.0 s consisted of a linear acceleration ramp up to this speed. These parameters were selected because they gave a uniform thickness distribution of the polymer film across the silicon wafer.

The samples were attached to a stainless steel sample holder held 45 mm beneath a stainless steel mesh to which the sample holder was electrically connected. This entire assembly was placed within the diffusion chamber of the plasma immersion ion implantation (PIII) system described in [3].

The plasma used for the ion implantation was an inductively coupled RF plasma powered at a frequency of 13.56 MHz. The plasma power was 100 W with a matched reverse power of 12 W. The base pressure of the diffusion chamber was $10^{-4}$ Pa with an argon pressure of $4.4 \times 10^{-2}$ Pa and a nitrogen pressure of $2.7 \times 10^{-2}$ Pa. Pulsed bias was applied to the sample holder and mesh assembly. For films with 100 nm thickness, the bias voltage was set at -20 kV and the pulses were applied at a frequency of 50 Hz for a total time of 400 sec. The 20-30 nm thick samples were treated using a bias voltage of -2 keV with pulse frequency of 1600 Hz. The bias voltage was reduced for the thinner films in order to constrain the ion collisions and recoils to a depth of 10 nm from the surface so that the ions would not disrupt the film-substrate interface appreciably. The modification time was reduced to prevent the etching of the film by the ambient plasma. The lengths of the applied bias pulses were 20 μSec. The parameters were chosen to achieve a fluence of $10^{16}$ ions $cm^{-2}$.

*2.2 Thin film characterization*

The samples were studied using a visible light ellipsometer in order to estimate the thickness and optical properties (refractive index and extinction coefficient) of the untreated and modified PS layers. A Woollam M2000V spectroscopic ellipsometer was used for the measurements. Ellipsometry data were collected for three angles of incidence: $65^0$, $70^0$ and $75^0$ in the wavelength range 400-800 nm. A model consisting of a Cauchy layer on top of a silicon substrate was used to fit the data. Thickness and optical constants associated with the best fit model were determined for the untreated and PIII treated PS films.

FTIR-spectroscopic ellipsometry measurements were performed in the infrared region with a J.A. Woollam Co. IR-VASE (Variable Angle Spectroscopic Ellipsometer) with integrated BOMEM FTLA 2000 FTIR spectrometer. The ellipsometer was used in a rotating compensator configuration. Ellipsometric psi and delta data with spectral resolution of 8 $cm^{-1}$ in the spectral range of 400–4000 $cm^{-1}$ (20 scans per spectrum, 20 spectra per one polarizer position, 30 turns of polarizer) were acquired at 50, 60 and 70° angles of incidence. These angles near the Breswter angle were selected to achieve good sensitivity. The ellipsometric modelling was done with the WVASE-32 software by fitting a multilayer model. To minimise the number of fitting parameters at each step, the model was built up layer by layer with new ellipsometric data taken with the addition of each new layer. The $\Psi$ and $\Delta$ spectra of the bare silicon wafer were well fitted by a model consisting of a substrate of silicon and a single layer of silicon oxide, using the known constants of silicon and silicon oxide. After deposition of the polystyrene coating and PIII modification, new ellipsometric data were taken and the PS was modelled as a Cauchy layer over the spectral range 2000-2500 $cm^{-1}$ which does not contain absorption lines to obtain the thickness. Using this thickness a model was fitted over the entire wavelength range.

FTIR transmission spectra from the samples were recorded using a BOMEM FTLA 2000 FTIR spectrometer. To obtain sufficient signal/noise ratio and resolution of spectral bands, we used 500 scans and a resolution of 4 $cm^{-1}$. The spectrum from a silicon wafer was subtracted. All spectra were baseline corrected.

Micro-Raman spectra ($\lambda$=532.14 nm) were obtained in the backscattering mode using a diffraction double monochromator spectrometer HR800, Jobin Yvon, LabRam System 010. The spectral resolution was 4 $cm^{-1}$ and the number of scans and integration time were varied to ensure sufficient signal-to-noise ratio. An optical microscope (Olympus BX40) with a 100× objective was used to focus the laser beam and collect the scattered light. LabRam$^{TM}$ software was used to analyze the spectra.

Atomic Force Microscopy (AFM) images of samples were collected on a Pico SPM instrument in tapping mode, at a scan rate of 0.5 lines/sec over areas of 5x5 and 1x1 μm, tip radius is 10 nm. Analysis of the AFM images was performed using the WSxM software (version 3, Nanotec Electronica S.L. Spain) [25].

X-ray reflectometry measurements of the samples were made at the Australian Nuclear Science and Technology Organization (ANSTO) in Sydney, Australia, using a Panalytical X'Pert Pro reflectometer and Cu K$\alpha$ radiation (wavelength 1.54056 Å). The specularly reflected beam was measured as a function of the momentum change perpendicular to the surface ($Q_z = 4\pi(\sin\theta)/\lambda$) over the range 0.007-0.7 $Å^{-1}$. The X-ray beam was generated with a 45 kV tube source, focused using a Göbel mirror and collimated with 0.2 mm pre- and post-sample slits, before being detected with a NaI scintillation detector. The reflectometer allowed for the resolution of subsurface scattering length density (SLD) variations throughout the thin films as a function of depth. The MOTOFIT software package [26] was used to determine the X-ray SLD profile for a given film, based on the Abéles formalism for planar multi-layers. Due to the lack of scattering contrast between Si (SLD = 2.01×$10^{-5}$ $Å^{-2}$) and the native oxide layer (SLD = 1.89×$10^{-5}$ $Å^{-2}$), the $SiO_2$ layer was omitted from structure refinements using X-ray reflectivity data.

Neutron reflectivity data were measured using the Platypus time-of-flight neutron reflectometer [27] and a cold neutron spectrum (2.8 Å $\leq \lambda \leq$ 18.0 Å) at the OPAL 20 MW research reactor. 23 Hz neutron pulses were generated using a disc chopper system (EADS

Astrium GmbH) in the medium resolution mode (ΔQ/Q = 4%), and recorded on a 2-dimensional helium-3 neutron detector (Denex GmbH). Reflected beam spectra were collected for each of the untreated and PIII modified PS films at 0.65° for 1 hour (0.5 mm slits), 2.0° for 3 hours (2 mm slits) and 4.0° for 5 hours (4 mm slits). Direct beam measurements were collected under the same collimation conditions for 1 hour each. Structural parameters for these films were also refined using the MOTOFIT reflectivity analysis software [26]. In the case of models refined using neutron reflectivity data, an additional ~10 Å layer was included adjacent to the silicon (SLD = $2.07\times10^{-6}$ Å$^{-2}$), representing the native oxide layer (SLD = $3.47\times10^{-6}$ Å$^{-2}$).

*2.3. Protein adsorption study*

After initial characterisations, the samples were incubated in a 10.0 μg/ml tropoelastin in PBS (phosphate buffered saline) solution at room temperature. The synthetic tropoelastin used in these experiments was SHELΔ26A, a 60.1 kDa isoform of human tropoelastin (lacking domain 26A), produced in the laboratory of Professor A. Weiss (University of Sydney) by over-expression in *E-coli* and purified according to established methods [28, 29]. The incubation time for saturation in protein coverage (55 minutes) was determined by measuring the layer thickness as a function of incubation time using visible light ellipsometry. After incubation the samples were rinsed in Milli-Q water and dried in atmosphere at room temperature and then measured. After measurement the samples were washed in 2.0% w/v SDS (sodium dodecyl sulfate) in water solution, at 80° C for 55 minutes and measured again and rinsed in Milli-Q water. The adsorbed tropoelastin layer before and after SDS elution was characterized using X-ray reflectometry, AFM, FTIR and ellipsometry. The strength of binding of the protein layer can be assessed by washing with detergent. SDS is a detergent that is used to unfold proteins and to remove physically adsorbed proteins from surfaces. The SDS cleaning procedure has been used as a method to test whether proteins are covalently attached to surfaces [30, 31].

**3. Results**

*3.1. Optical spectroscopy*

Raman spectroscopy shows that the PS coating is transformed into a disordered graphitic structure with PIII treatment. After PIII treatment with Ar ions of 20 keV energy and $10^{16}$ ions/cm$^2$ fluence, the Raman spectra change dramatically: a strong broad band, as shown in Figure 1, appears and the polystyrene lines observed for the untreated film disappear. The new band is typical of graphite-like amorphous carbon. The band can be fitted by the G-peak at 1543 cm$^{-1}$ and the D-peak at 1384 cm$^{-1}$ characteristic of disordered carbon structures containing a majority of sp$^2$ bonds [32].

Figure 2 shows the FTIR spectral changes associated with (a) the PIII treatment of the PS coating; (b) the surface adsorption of a tropoelastin layer onto the PIII treated surface; and (c) washing of the same tropoelastin layer in SDS detergent. The spectrum of the PIII modified PS shows vibrational lines of residual hydrocarbon groups in the range 2800-3000 cm$^{-1}$ attributed to CH stretching, $\nu$(CH), and at 1450 cm$^{-1}$ attributed to bending vibrations $\delta$(C-H), as well as lines from unsaturated carbon-carbon group vibrations, $\nu$(C=C), in the 1600-1650 cm$^{-1}$ region. Absorptions associated with vibrations of oxygen-containing groups at 3500 cm$^{-1}$, $\nu$(OH), attributed to hydroxyl, peroxide, carboxyl groups and at 1720 cm$^{-1}$, $\nu$(C=O), attributed to carbonyl, carboxyl and ester groups, are observed.

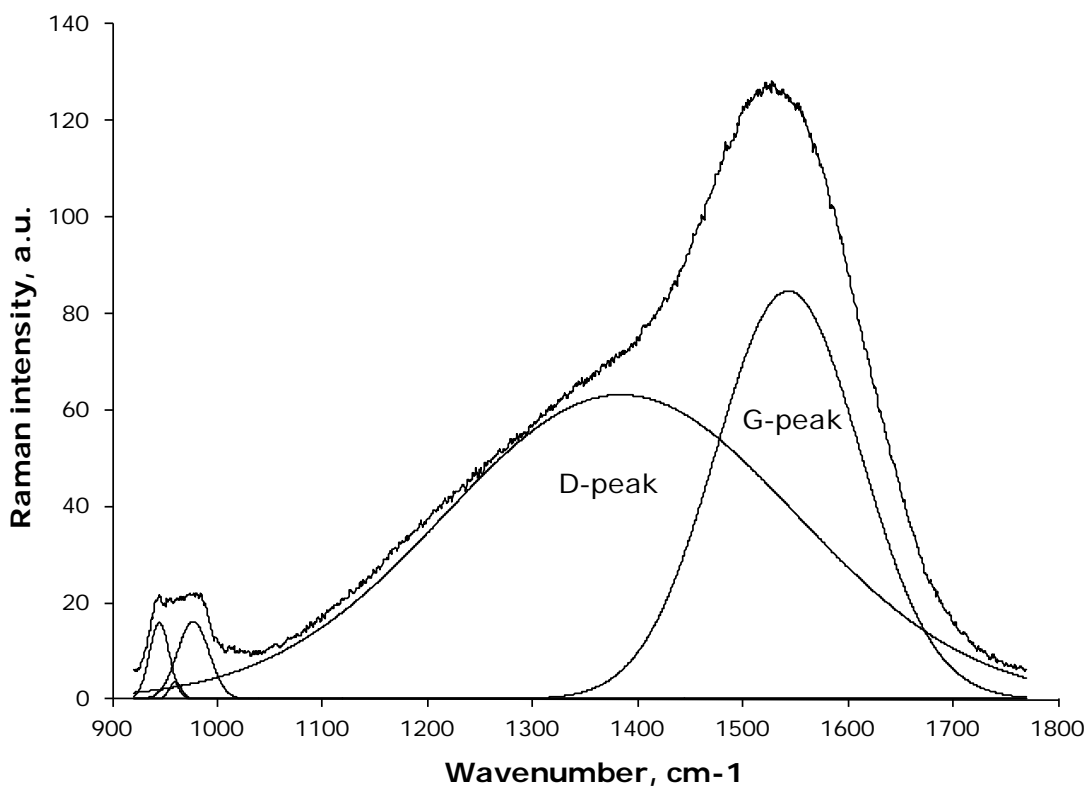

**Figure 1.** Micro-Raman spectra and peak fitting for a PS coating on silicon wafer after plasma immersion ion implantation with nitrogen using -20 kV pulse bias.

The spectra of the modified PS layer with attached tropoelastin show new lines, corresponding to the Amide A, I and II at 3300, 1650 and 1540 cm$^{-1}$ vibrations, characteristic of protein molecules. After washing of the sample in SDS detergent, the protein lines remain in the spectra with reduced intensity, showing that a significant fraction of the protein is covalently bound to the surface. In the case of untreated polystyrene no protein lines remain after SDS washing indicating that all of the tropoelastin has been removed.

Additional lines at 1720, 1580 and 1450 cm$^{-1}$ appear, corresponding to groups produced by chemical reaction with SDS. Sodium reacts with carboxyl groups to form sodium carboxylate (1580 cm$^{-1}$ line attributed to the $\nu$(C=O) vibrations in the sodium carboxylate group) causing the surface layer to swell. The residual acid part of SDS reacts with hydroxyl groups to form esters (1720 cm$^{-1}$ line). The presence of hydrocarbon chains fixed in the surface layer is observed by the appearance of new bending $\delta$(CH$_2$) vibrations (1450 cm$^{-1}$ line). We observed the same reaction products in the surface modified layer in the XPS spectra of PIII modified Nylon after incubation in buffer solution [33].

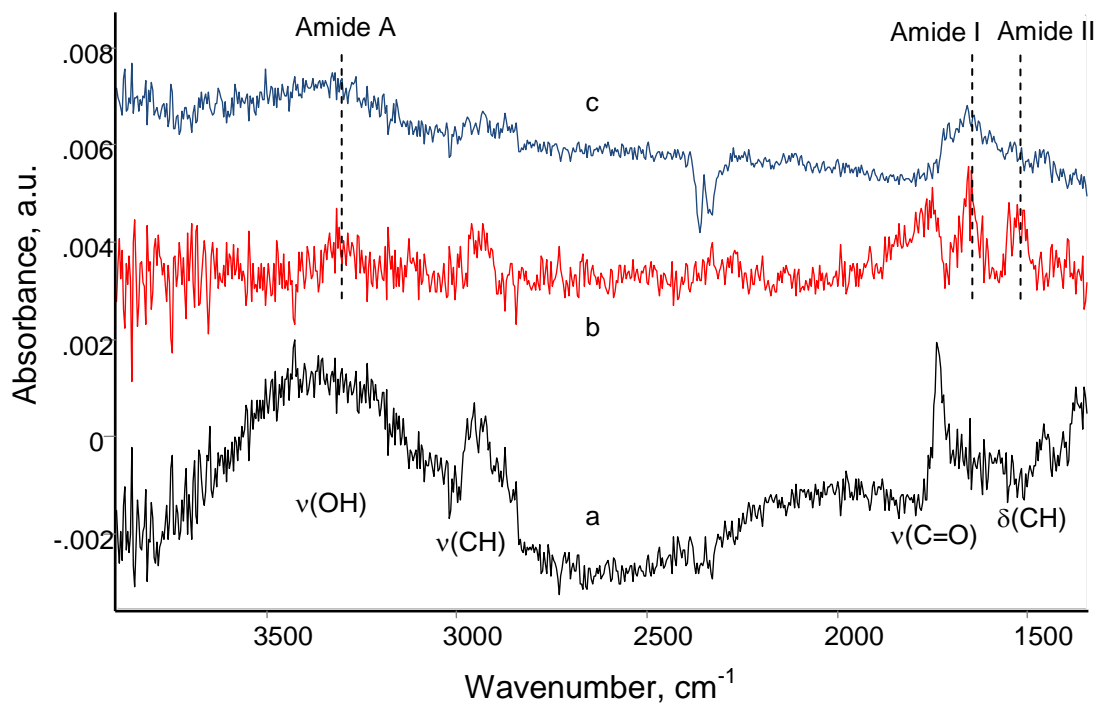

**Figure 2.** FTIR spectra of (a) 13 nm PS modified by PIII the arrows indicate the lines of groups introduced by the PIII modification and residual hydrocarbon structures; (b) the sample after incubation in tropoelastin with the spectrum of the PIII modified PS (a) subtracted; (c) the sample after incubation in tropoelastin and subsequent SDS washing with the spectrum of the PIII modified PS (a) subtracted. The lines due to the protein vibrations Amide A, I and II are indicated. All spectra are baseline corrected and have the spectrum of the uncoated silicon wafer subtracted.

*3.2. Atomic Force Microscopy*

Figure 3 shows AFM measurements of the tropoelastin layers on untreated polystyrene indicating that the tropoelastin forms an incomplete monolayer and that the size of the tropoelastin molecule is about 6.4 nm in at least one dimension. The AFM topography images confirm that the dry tropoelastin layer contains large pits. Differences in tip interactions with the surface suggest that the low regions are PS while the high regions are protein. The roughness histogram shows the distribution of surface pixel events has two peaks (Figure 3(e, f)). The 6.4 nm distance between the peaks corresponds to the average distance between the PS surface and the top surface of the tropoelastin layer. To obtain an estimate for the protein coverage the AFM image was analyzed to obtain the fractional coverage of pits >6.4 nm in depth, and indicates that tropoelastin covers ~72% of the surface. The average pit size is in the range of 150-170 nm. Similar protein islands have also been previously observed for the enzyme horseradish peroxidase on untreated polystyrene [3].

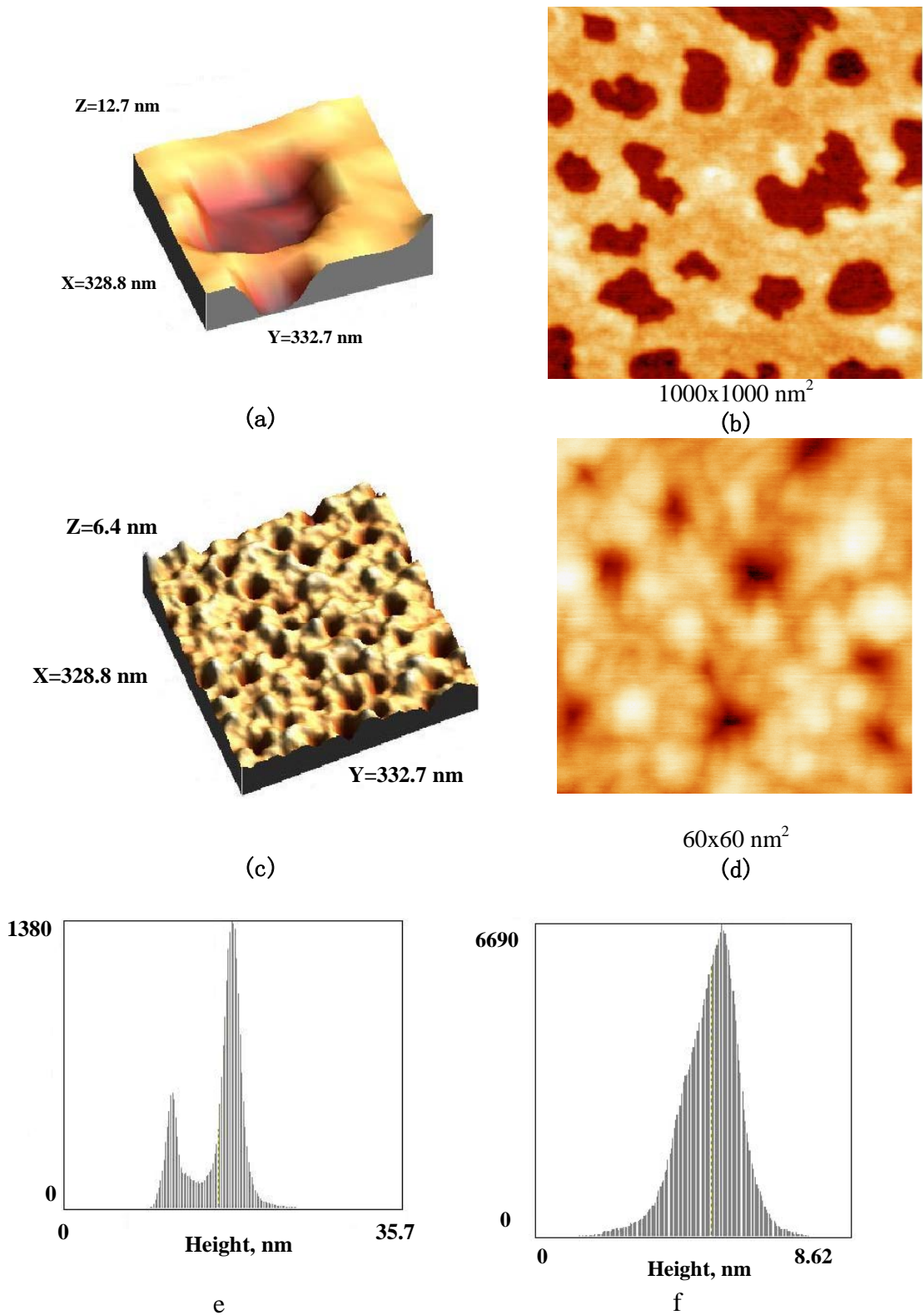

**Figure 3.** AFM topography image of tropoelastin attached on untreated PS (a, b) and on PIII treated PS with Ar ions of 20 keV energy and $10^{16}$ ions/cm$^2$ fluence (c, d). Image size is 328.8x332.7 nm (a), 1000x1000 nm (b), 328.8x332.7 nm (c), and 60x60 nm (d). Distribution of pixel events in area of 5x5 μm on height for AFM image of tropoelastin on untreated (e) and PIII treated PS (f).

The tropoelastin coverage on PIII modified PS does not have pits deep enough to be attributed to gaps in the protein coverage. The AFM phase images show that the material in low and high regions is similar. The distribution of surface pixel events in this case only has one peak, and the few pits in the surface that were observed have a depth of about 2.7 nm and narrow bases. The AFM image was analyzed to obtain the fractional coverage of pits >2.6 nm in depth, and indicated that tropoelastin covered 95% of the surface. The size of these pits is too small to allow for penetration of the protein molecules (Figure 3(c)).

*3.3. Ellipsometry*

Figure 4 shows optical constants of both treated and untreated PS films. For the untreated film, n is in the range 1.5-1.55 and k=0 with the measured thickness being equal to the thickness of the spin coated PS layer. After PIII treatment, the thickness of the PS film decreased due to plasma and ion etching (see Table 1). As a result of PIII induced graphitisation and densification, the value of n for the PS film increases to 1.65-1.72 and k is in the range 0.02-0.07 over the wavelengths fitted. The model is not sensitive to gradients in the structure of the modified layer since uniform and spatially varying profiles gave similar fits to the experimental results.

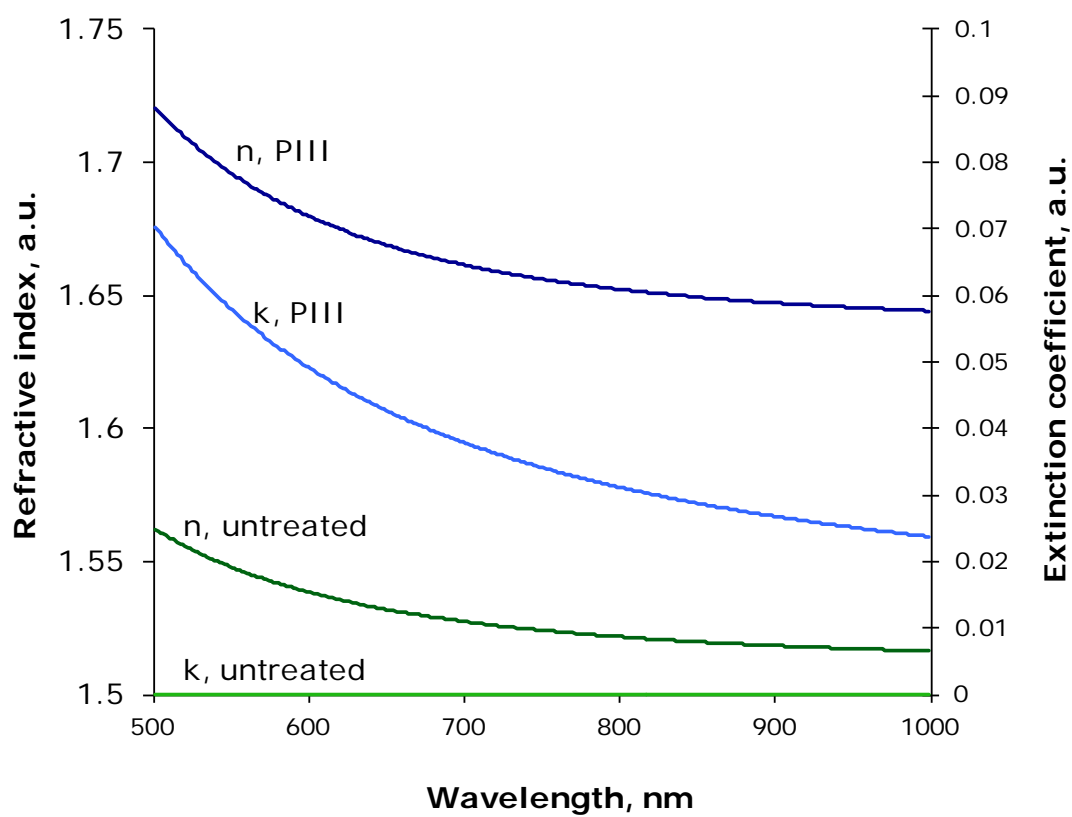

**Figure 4.** Refractive index and extinction coefficient of PS layer untreated and after PIII treatment with Ar ions of 2 keV energy and $10^{16}$ ions/cm$^2$ fluence.

The refractive index of a PIII modified PS layer measured using a FTIR ellipsometer (2500-25000 nm wavelength) is shown in Figure 5. The refractive index of the PIII modified PS coating is 2.1-2.3 in the 1200-4000 $cm^{-1}$ region. This refractive index indicates a high degree of carbonization of the PS by PIII treatment, as also observed by the FTIR transmission spectroscopy presented above. The fitting gave a thickness of 12.4 nm for the PIII modified PS coating. The peaks in the region of 1100-1200 $cm^{-1}$ correspond to the Si-Si and Si-O vibrations in the silicon substrate and which were not fully compensated.

After incubation in buffer solution and drying, the refractive index changed slightly (Figure 5) and the thickness increased to 13 nm. This is consistent with swelling and reactions with the buffer solution as discussed above.

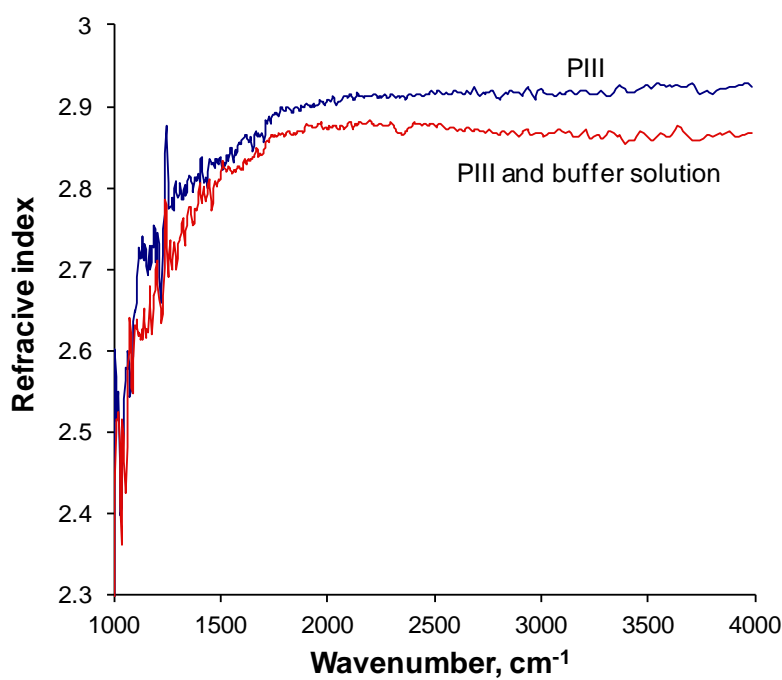

**Figure 5.** Refractive index of the PS layer after PIII treatment (blue) and after incubation in buffer solution (red). The small decrease of index after incubation is interpreted as a result of changes in chemical composition and swelling due to exposure to buffer solution.

After incubation in protein solution and drying, further ellipsometric data were acquired and modelled. The model consisted of the silicon substrate, a silicon oxide layer and the PIII modified PS layer after soaking in buffer, and a new layer representing the protein. The parameters for all but the new protein layer were obtained from the model fitted to the buffer incubated sample (Figure.5). A Cauchy layer was added to the model to account for the protein at the top. The thickness of protein layer was found to be about 4.9 nm, by fitting of spectra in 2000-2500 $cm^{-1}$ spectral range. Fixing the 4.9 nm thickness in the model, the optical constants, n and k, were then fitted over the whole spectral range to give the optical constants, including absorption features, for the protein layer. The optical constants of the protein layer show the two most intense lines of absorbance in the FTIR spectra of proteins: Amide I and Amide II (Figure.6). The amide A line (at about 3300 cm$^{-1}$) has a lower intensity and is not as well resolved from the noise signal.

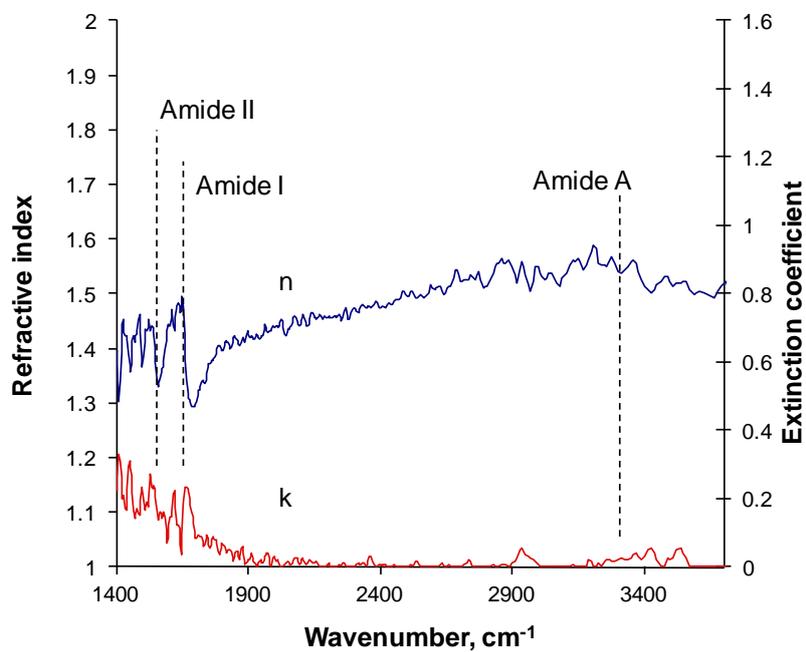

a

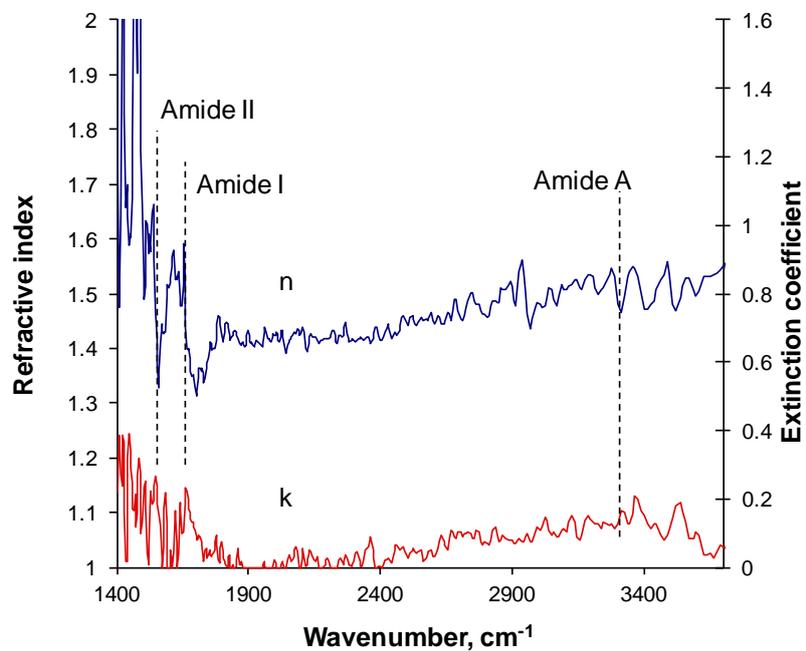

b

**Figure 6.** Refractive index (n) and absorption coefficient (k) for the PIII modified PS layer after incubation in protein containing solution, (a) before and (b) after washing in SDS detergent. The amide features are indicated by vertical lines and remain after washing.

The same fitting procedure was carried out for ellipsometric data obtained after the sample was washed in SDS detergent. The fitted optical constants of the protein layer were similar to that found prior to the SDS wash, while the thickness of the layer decreased to about 3.0 nm. The results of the fitting are presented in Table 1.

**Table 1.** Thicknesses of the polymer and adsorbed tropoelastin layers determined using FTIR-ellipsometry. The thickness of the protein layer was also determined with AFM and X-ray reflectivity methods. Note that the AFM measurement indicates the peak height measured whereas the other measurements give an average value of the thickness. The error for the AFM value was calculated from the width of the height event distribution.

| Layers | Spin Coating | PIII | Buffer Soaking | Protein Soaking | SDS Washing |
|---|---|---|---|---|---|
| Silicon | 1 mm | 1 mm | 1 mm | 1 mm | 1 mm |
| $SiO_2$ | 1 nm | 1 nm | 1 nm | 1 nm | 1 nm |
| PS | 16 nm | | | | |
| PIII PS | | 12.4 nm | 13.0 nm | 13.0 nm | 13.0 nm |
| Protein (ellipsometry) | | | | 4.9(5) nm | 3.0(5) nm |
| Protein layer (AFM) | | | | 6.4(5) nm | |
| Protein layer (X-ray reflectomery) | | | | 2.4(1) nm | 2 nm |

*3.4. X-ray and Neutron Reflectometry of PS and protein films*

In each case described below, attempts were made to fit the observed X-ray reflectivity (XRR) data using single layer models. Such a model proved to be appropriate only in the case of virgin PS films prior to PIII modification. Figure 7 shows X-ray reflectivity data from (a) an untreated film, and argon ion treated PIII modified films (2 kV, 1600 Hz) after (b) 80 s and (c) 400 s. The points are the observed data while the solid line is calculated reflectivity based-on the structural models described below. By following the spacing between minima in the Kiessig fringes, one can see that the thicknesses of these films are decreasing with increasing treatment time due to etching of material from the surface. In the case of the untreated polystyrene film (Figure 7(a)), a very good fit was obtained using a single-layer model. The thickness of this film was determined as 26.4(1) nm and a constant SLD of $9.4(1)\times10^{-6}$ $Å^{-2}$ (Figure 8(a)). This refined SLD equates to a film mass density of 1.03 $g/cm^3$ and is typical of PS thin films prepared by spin coating. The Si/film interfacial roughness was refined as 0.3(1) Å, while the surface roughness of the PS film was 0.4(1) Å.

In the case of X-ray reflectivity data for the PIII modified films (Figures 11(b) and 11(c)), attempts to fit these data to a single layer models were unsuccessful, and so a different approach was used to allow fitting of films having variable composition and density. The final analysis of XRR data from these argon ion PIII modified films was made using the Discrete Density Profile (DDP) method described by Fullagar and co-workers [34]. In this fitting method, the number, thickness and roughness of layers are chosen for the structural model based on the quality of the data (i.e. the value of $Q_{max}$ where the specular reflectivity is no longer discernable above the background). With the exception of the uppermost layer, the thickness and interfacial roughness of the layers were fixed, and only the Scattering Length Density of each layer was refined to generate a smoothly varying SLD profile. For the refinement of the XRR data in Figure 8(b) 12 layers were used, and for Figure 8(c) 10 layers were used; with the thickness of each layer was fixed at 2 nm, and the interfacial roughness was fixed at 0.4 nm. An excellent fit to these data were obtained in each case, with SLD profiles for the PIII modified films of thickness 24.2(1) nm (80 s) and 20.0(1) nm (400 s) shown in Figures 8(b) and 8(c) respectively.

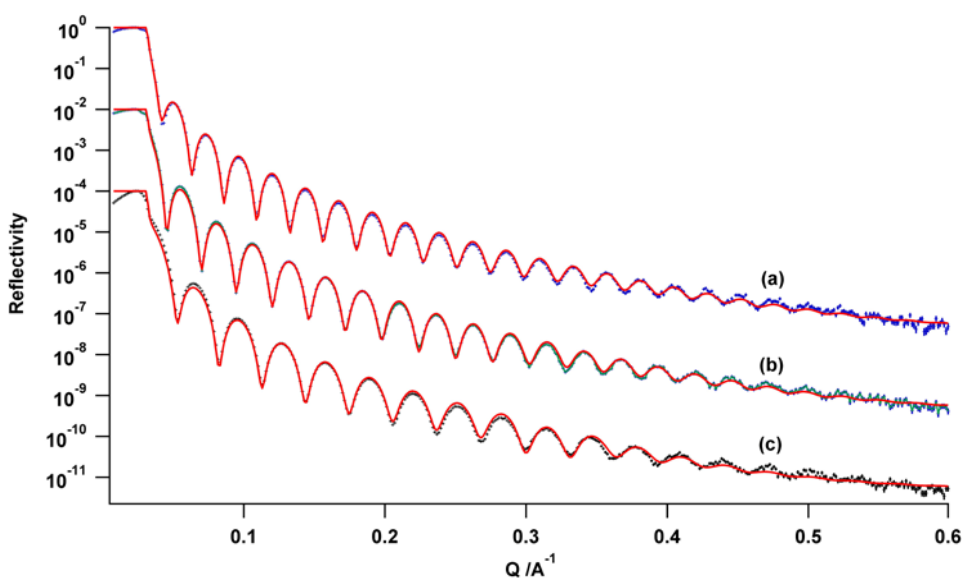

**Figure 7.** Observed (points) and calculated (red curve) X-ray reflectivity from **(a)** (blue data) untreated spin coated PS film; **(b)** (green data, offset by $10^{-2}$ for clarity) PIII treated PS (argon ions, 2 kV bias, treatment 80 s); and **(c)** (black data, offset by $10^{-4}$ for clarity) PIII treated PS (argon ions, 2 kV bias, treatment 400 s).

The above set of films was also characterized before and after argon ion PIII modification using neutron reflectometry. While both X-ray and neutron reflectivity probes the film thickness and internal structure, the former is sensitive to the electron density of the film and the latter is sensitive to the neutron scattering length density. In each case, the X-ray and neutron SLDs are a combination of the composition and mass density at any level within the film [35]; although, neutron reflectometry is much more sensitive to hydrogen content than XRR due to the substantial difference in coherent neutron scattering length (b) between C (6.646 fm = $10^{-15}$ m) and H (-3.739 fm).

Figure 9 shows the equivalent observed (data points) and calculated (solid lines) neutron reflectivity data collected using the Platypus reflectometer at the OPAL research reactor for (a) untreated, and argon ion treated PIII modified films (2 kV, 1600 Hz) after (b) 80 s and (c) 400 s. Again, data from the untreated PS film were fitted using a single-layer model; while data from PIII modified films were modeled using the DDP method. Refined neutron SLD profiles based on these data are shown in Figure 10.

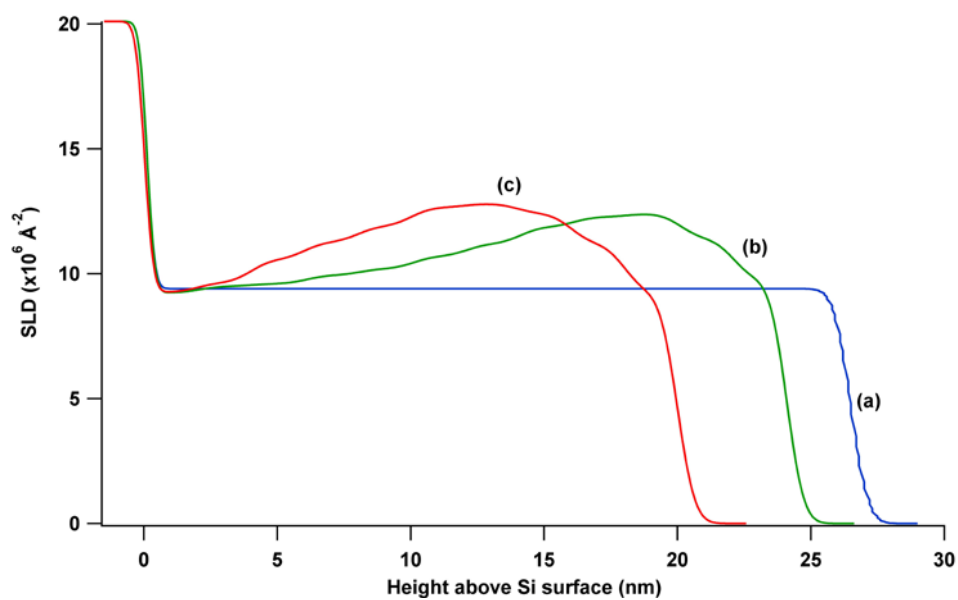

**Figure 8**. X-ray SLD profiles from **(a)** (blue curve) untreated spin coated PS film; **(b)** (green curve) PIII treated PS (argon ions, 2 kV bias, treatment 80 s); and **(c)** (red curve) PIII treated PS (argon ions, 2 kV bias, treatment 400 s).

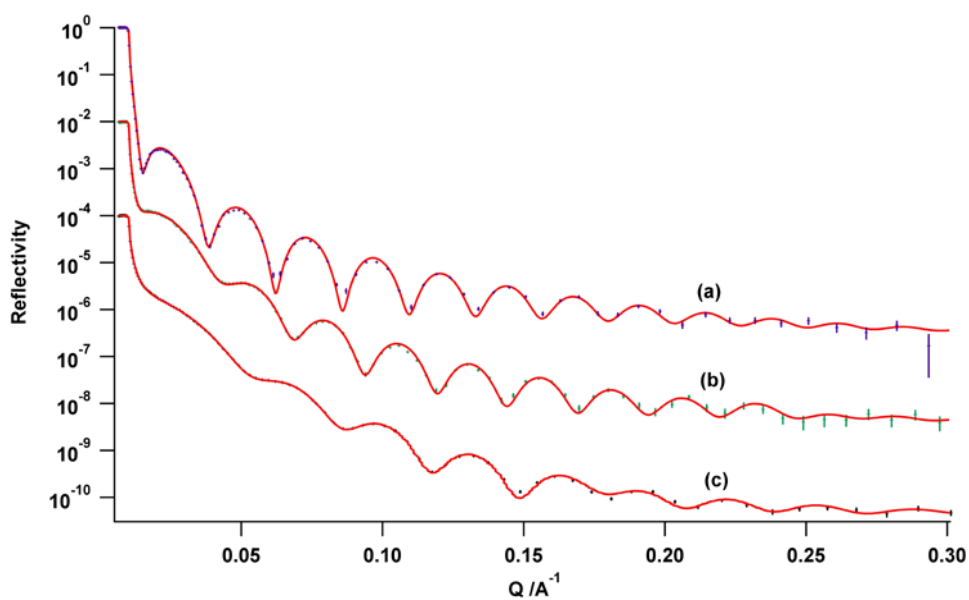

**Figure 9.** Observed (points) and calculated (red curve) neutron reflectivity from **(a)** (blue data) untreated spin coated PS film; **(b)** (green data, offset by $10^{-2}$ for clarity) PIII treated PS (argon ions, 2 kV bias, treatment 80 s); and **(c)** (black data, offset by $10^{-4}$ for clarity) PIII treated PS (argon ions, 2 kV bias, treatment 400 s).

X-ray reflectivity was used to analyze the attached protein on the untreated and PIII treated PS. First, the measurement was done for a ~14 nm as-prepared PS coating, before and after incubation in tropoelastin solution. In each case, attempts were made to fit X-ray reflectivity

(XRR) data using single layer model. Such a model proved to be appropriate only in the case of virgin PS films before the incubation, with the thickness determined to be 14.2(1) nm and a constant SLD of $9.5(1) \times 10^{-6}$ Å$^{-2}$ and a surface roughness of 0.3(1) nm. A single layer model did not lead to a satisfactory fit to the data of the protein incubated film ($\chi^2 = 0.0091$); although a 2-layer model gave a substantially better fit to these data ($\chi^2 = 0.0045$), leading to a base layer of polystyrene (of 14.1(1) nm thickness and SLD $9.0(1) \times 10^{-6}$ Å$^{-2}$) and a 2.0(1) nm thick layer of increased SLD (SLD $9.4(1) \times 10^{-6}$ Å$^{-2}$) at the surface of the film. Based on an average molecular formula for the protein of ($C_{2747}H_{4405}N_{759}O_{773}S_2$) the mass density of the adsorbed layer is estimated as 1.03 g/cm$^3$. The density of protein is in a range of 1.33-1.42 g/cm$^3$. An average value for the density of protein is 1.35 g/cm$^3$ [36]. The low density, we observe, must be due to incomplete coverage of the PS surface. The calculated coverage of PS surface from the X-ray fitted data is 76% and agrees with the AFM measurement of 72%.

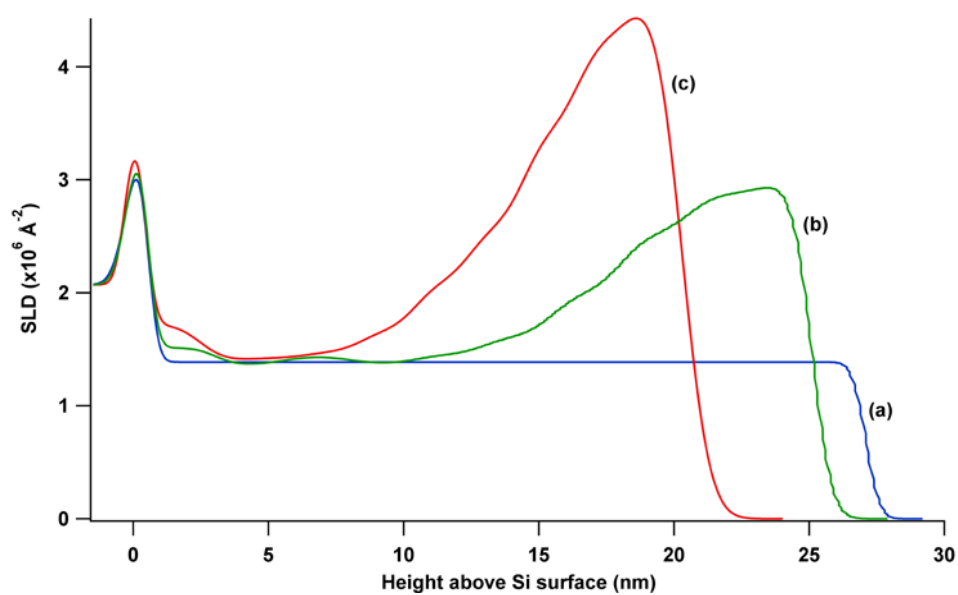

**Figure 10**. Refined neutron SLD profiles from **(a)** (blue curve) untreated spin coated PS film; **(b)** (green curve) PIII treated PS (argon ions, 2 kV bias, treatment 80 s); **(c)** (red curve) PIII treated PS (argon ions, 2 kV bias, treatment 400 s). -3.74 H, 6.65 C, 5.80 O, 4.1 Si

Next, the PS film modified by PIII treatment of 2 kV argon ions for 400 s was used for protein attachment. A very good fit of the X-ray reflectivity data generated using the DDP method from modified PS film was obtained via the DDP method ($\chi^2 = 0.005$), using 11 layers, giving a total film thickness of 9.9(1) nm and a surface roughness of 0.4(1) nm. The attachment of protein changes the observed X-ray reflectivity, leading to more closely spaced Keissig fringes and indicating an increase in total film thickness. Fitting using the DDP method gave a very good fit ($\chi^2 = 0.005$) and in this instance required 13 layers of total thickness 12.3(1) nm, suggesting a thickness of the protein layer of approximately 2.4 nm. The 2-3 nm surface layer of this film clearly reveals a monolayer of adsorbed tropoelastin protein, having a slightly lower refined SLD of $\sim 12 \times 10^{-6}$ Å$^{-2}$ than the graphitized bulk of the PIII film. Although of similar thickness, the tropoelastin layer on this modified surface (1.32 g/cm$^3$) has a significantly higher density than the equivalent layer on the untreated PS film.

Using a value for the average density of protein of 1.35 gives a coverage of 98% in good agreement with the AFM measurement of 95%.

Subsequent washing with SDS detergent changes the X-ray reflectivity profile yet again with a good fit to these data ($\chi^2$ = 0.005) generated by an 11 layer DDP model. The overall thickness of the film was smaller, although it had increased slightly compared to the film before incubation with tropoelastin solution. We believe that this increase in SLD is due to the chemical modification of the modified PS layer by SDS. The presence of carboxyl groups in the modified PS layer, makes a reaction of sodium with carboxylic acid groups likely (see (1)). The products of this reaction are observed in the FTIR spectra as a broad peak at 1580 cm$^{-1}$ corresponding to sodium carboxylate vibrations, as discussed above.

## 4. Discussion

*4.1. The Internal Structures of PIII Modified Polystyrene Films*

Comparison between X-ray and neutron refined Scattering Length Density profiles (eg for argon ion treated polystyrene, Figures 8 and 10) clearly show a decrease in total film thickness due to plasma etching with increasing treatment time. Both X-ray and neutron reflectometry measurements of PIII modified polystyrene coatings indicate the presence of SLD gradients within these films. In all cases, the topmost layer has a SLD higher than the unmodified material that can be attributed to the carbonized structure with a higher mass density, inferred also from the Raman spectra and ellipsometry. The layer adjacent to the silicon wafer has a low SLD and mass density similar to the initial polystyrene coating. By inference, this layer contains the residual aliphatic groups and the oxygen-containing groups identified in the infrared spectra, since it is unlikely these groups are present in the carbonized layer.

Comparison between Figures 8 and 10 however shows quite distinct behaviors due to the fact that X-rays are sensitive to variation in electron density, while neutrons are particularly sensitive to hydrogen content. This allows us to comment on the relative changes in composition and mass density within these films. Unlike previous studies of RF plasma polymerized coatings of uniform composition [16, 17], the combination of these techniques in our study does not allow the decoupling of mass density and composition in these graded PIII modified films. X-ray reflectivity indicates a maximum in electron density approximately 5 nm below the surface for films after both 80 s and 400 s treatment, and a slightly higher overall SLD for the latter. In contrast, the neutron SLD has a maximum value for these modified films essentially at the surface; indicating a decrease in hydrogen content, coupled with increasing mass density as the surface is approached. The peak in neutron SLD for the film after 400 s (4.4×10$^{-6}$ Å$^{-2}$) is substantially larger than the film after 80 s (2.9×10$^{-6}$ Å$^{-2}$). By itself, this result could indicate a lower hydrogen content or higher mass density; however when coupled with similar X-ray scattering length densities it points to the former. When compared to the expected neutron SLD for graphite however (7.4×10$^{-6}$ Å$^{-2}$), the refined values suggest a residual hydrogen content in this part of the film, or a substantially lower mass density.

SRIM code calculation of argon ion penetration supported the observed structure defects in PS layer. The PIII treatment changes the polystyrene coating dramatically. The penetrating high energy ions cause collision cascades with atoms of the polystyrene macromolecules. As a result, some carbon and hydrogen atoms receive substantial energy and are move more deeply in to the structure. The distributions of knock-on hydrogen and carbon atoms and its new distribution as calculated using the SRIM code [37] are presented in Figure 11. The calculations were carried out with argon ions of 2 keV. Some hydrogen atoms may be captured by neighboring atoms in the new environment, but the majority diffuses outwards to the surface and leave the coating. The carbon atoms and ions move deeper into the modified layer (see shift between "carbon vacancy" and "carbon new positions" curves in Fig.11). This results in the formation of carbonized structures beneath the surface and the formation of pores structure in top of surface. The whole coating becomes increasing carbonized as the fluence of ions increases.

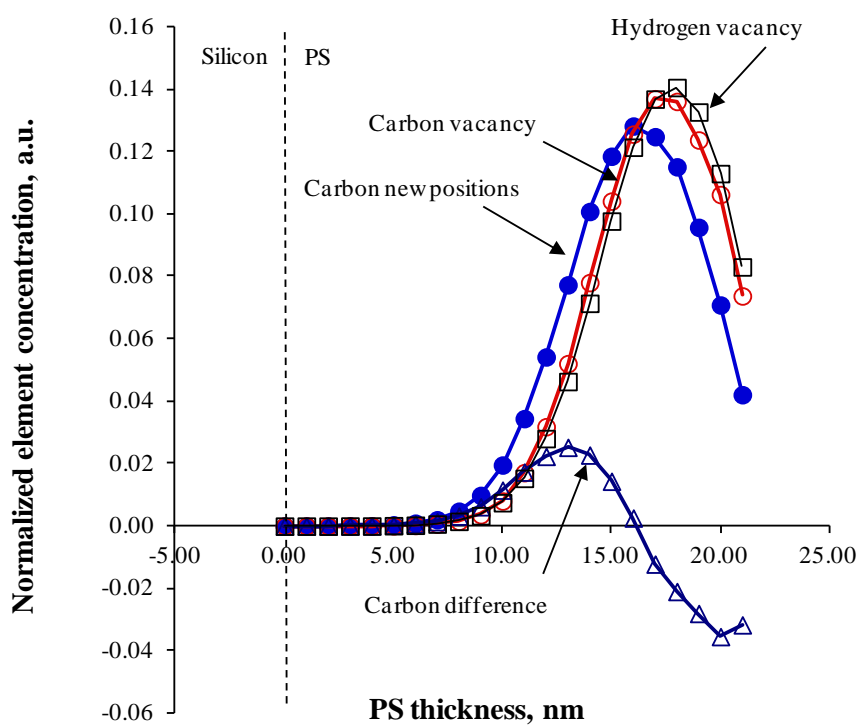

**Figure 11.** A SRIM simulation showing the results of implanting 2 keV argon ions into a PS target. Plotted are the depth distributions for hydrogen vacancies (squares), carbon vacancies (empty circles), displaced carbon atoms (solid circles), change in concentration of carbon atoms (triangles). The concentration of vacancies is normalized by the area under the curve for each curve. The region of 18-22 nm has carbon deficit. The region of 8-17 is carbon rich. The region of 10-22 nm has hydrogen deficit.

Taking into account SRIM calculation results, the SLD profiles of XRR and NR were used for calculation of chemical profile of the film. A chemical structure of the film was assumed to be $C_zO_yH_x$, where the concentration of oxygen was calculated as 10% following XPS data. Because of a low sensitivity of SLD function to oxygen concentration in comparison with hydrogen concentration, the oxygen profile was assumed constant over the film. The profile of hydrogen and density of the film were calculated to fit SLD profiles of XRR and NR. The deviation of 5% value for SLD of XRR and 10% for SLD of NR was used. The results of calculations are shown on Fig.12 and Table 2.

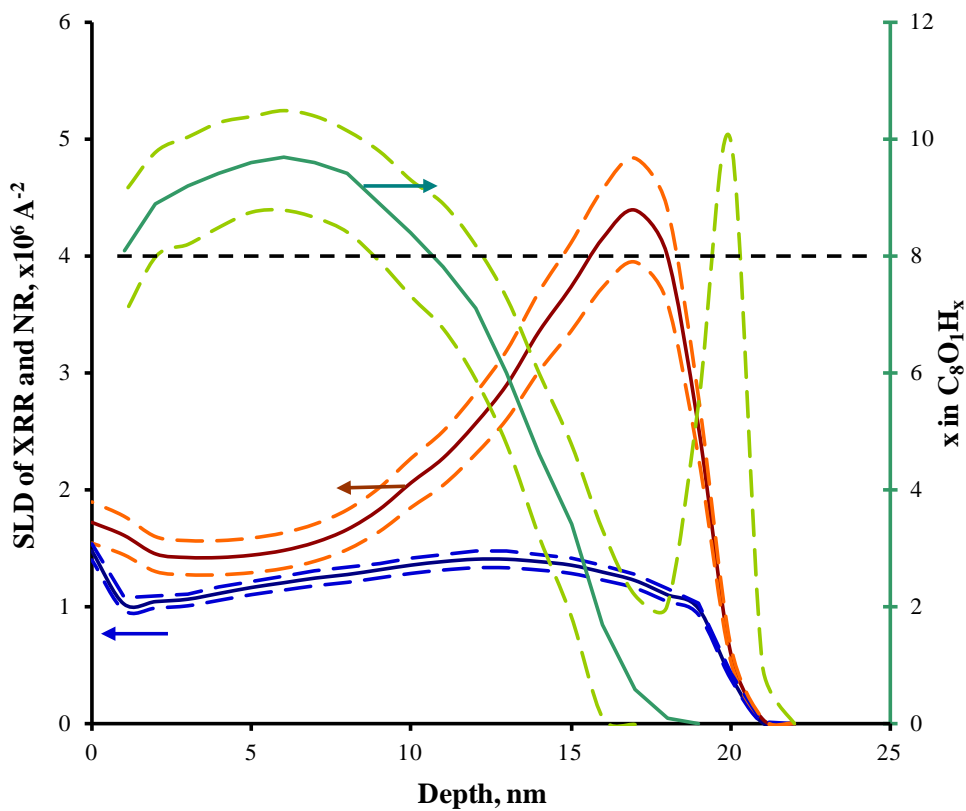

Figure 12. Calculated hydrogen content from SLD profile of XRR and NR data with error deviations (dashed lines). The sample of PS treated 400 sec of PIII time. Parallel dashed line at x=8 corresponds to untreated PS. The region of 2-10 nm has higher concentration of hydrogen than untreated PS. The region of 10-18 nm is dehydrogenated. The region of 18-22 nm has non-uniform structure.

Following fitted data, the total mass of the coating increases with PIII treatment (Table 2). The addition of the mass is due to oxidation of the coating. Due to decrease of the thickness and mass increase, the density of the coating increases with PIII treatment time. The density profile of the coating after 400 sec of PIII treatment is between initial PS density (1.03 g/cm3) and amorphous carbon density (1.8-2.1 g/cm3). It means that, the coating is not etched, but packed densely and collect oxygen at PIII treatment.

Table 2. Average mass and density of the PS coating before and after PIII, ratio oxygen/carbon from XPS spectra and proposed formula of the coating.

|  | untreated | 80 sec | 400 sec |
|---|---|---|---|
| mass, g/cm2 | 2.58E-06 | 2.94E-06 | 2.86E-06 |
| density, g/cm3 | 1.03 | 1.18 | 1.24 |
| O/C ratio | 0 | 0.125 | 0.225 |
| Formula | C8H8 | C8H7.2O1 | C8H7.1O2.3 |

The profile of hydrogen has maximum at 5-7 nm. The hydrogen content is this region is higher that in initial untreated polystyrene. The elevated value is higher that available error deviation. The high hydrogen content is caused by trapping of recoiled hydrogen atoms from the top surface layer by unmodified polystyrene molecules with formation of methyl groups. The appearance of methyl groups is observed by FTIR transmission spectra.

The low hydrogen content is observed in top surface layer that goes to zero in 18 nm region of the film. In this layer the hydrogen completely removed from the structure. The fitting of 18-22 nm region is unstable. The hydrogen profile shows high deviation up to negative values, that does not have physical meaning. This region can be interpreted as highly disordered with high fluctuations of density. Similar disordered structure is observed in results of molecular dynamic modeling of the polystyrene bombarded by high energy ions.

*4.2. Enhanced Adsorption of Protein on PIII Modified Surfaces*

A surface attached protein layer is clearly detected after incubation in tropoelastin solution by FTIR transmission spectroscopy, FTIR ellipsometry and X-ray reflectometry. The presence of a polypeptide backbone in this layer is revealed by characteristic amide group vibrations in the FTIR transmission and FTIR ellipsometry spectra. The FTIR ellipsometry data (4.9 nm) and AFM (6.4 nm) indicated a thicker protein layer compared to the values obtained by X-ray reflectivity (2.4(1) nm). While small angle X-ray scattering has been reported on a peptide fragment (SHEL 21–23N) of tropoelastin (indicating for example a radius of gyration of 1 nm and a length of 4.2 nm) [38]; to date there has been little published information on the dimensions of surface-bound tropoelastin protein. A recent study by Chow [39] using X-ray reflectometry and AFM, revealed SHELΔ26A tropoelastin monolayers of thickness 4.8 nm and 6.1 nm, deposited on hydrophobic octadecyltrichlorosilane self-assembled monolayers following incubation in water and PBS respectively. The tropoelastin monolayer on unmodified PS referred by J. Holst at al [40] is 6.4 nm. The latter value in particular, is essentially the same as the value that we have determined using AFM for tropoelastin adsorbed from PBS on hydrophobic untreated polystyrene.

The SLD obtained from X-ray reflectivity corresponds to the expected value for a protein layer. The SLD profiles strongly suggest that the protein molecules sit on top of the coating without penetrating it, and AFM measurements support this interpretation. These two techniques also paint a consistent picture when comparing tropoelastin coverage for untreated and PIII modified polystyrene films. Analysis of AFM data shows an increase in surface coverage from 72% to 95% of the surface; while X-ray reflectometry indicates an increase in density from 1.03 g/cm$^3$ to 1.32 g/cm$^3$ corresponding the incomplete protein coating of 76% and 98%.

After washing with SDS detergent, FTIR-ellipsometry and X-ray reflectivity data indicate that the protein layer becomes thinner, but that a residual amount of protein remains on the surface. This supports previous results that propose that the binding of up to a monolayer of protein to ion beam modified polymer surfaces can be covalent [8-10].

5. Conclusion

The combination of Raman, FTIR transmission spectroscopy, FTIR ellipsometry, X-ray and neutron reflectivity measurements together have been used to establish a model for the structure of polystyrene after PIII modification. In this model, the ion beam modification develops a graded structure in which the near surface layers are highly dehydrated with porous carbon structure, underneath layer is highly carbonized with high density, and the bulk layer is saturated hydrocarbons. The film is oxidized. The coating has a high refractive index and a high SLD for X-ray and neutrons. The mass of the film is higher than before treatment due to oxygen incorporation. The decrease of film thickness is caused by densification of the film, but not etching. After incubation of the modified polystyrene in tropoelastin solution our analyses establish that a protein layer binds to the surface. Detergent washing removes this layer only partially indicating that some of the protein is covalently bound to the modified surface.

Acknowledgments

We would like to gratefully acknowledge the Australian Research Council and AINSE for supporting this project.